\documentclass[vecphys]{svmult}

\usepackage{makeidx}         
\usepackage{graphicx}        
\usepackage{multicol}        
\usepackage[bottom]{footmisc}

\makeindex             

\begin{document}

\title*{First results of galactic observations with MAGIC}

\author{N\'uria Sidro\inst{1}
for the MAGIC Collaboration\inst{2}}
\institute{Institut de F\'isica d'Altes Energies (IFAE), Universitat Aut\`onoma de Barcelona, 08193, Bellaterra, Spain
\texttt {nuria.sidro@ifae.es}
\and Updated collaborator list at \texttt{http://wwwmagic.mppmu.mpg.de/collaboration/members}}
%
%
\maketitle
\index{N. Sidro}

\abstract{
During its first cycle, the MAGIC (Major Atmospheric Gamma-ray Imaging
Cherenkov) telescope was performing an observational campaign covering
a total of about 250 hours on galactic sources. Here we review the results
for the very high energy ($>100$ GeV) $\gamma$-ray emission from some of
those sources. 
}
\section{Introduction: The MAGIC Telescope} 
\label{sec:intro}
The Major Atmospheric Gamma Imaging Cherenkov 
(MAGIC) telescope~\cite{performance} is a very high energy (VHE) $\gamma$-ray
telescope, operating in an energy band from 100 GeV to 10 TeV,
exploiting the Imaging Air Cherenkov technique. Located on the
Canary Island of La Palma, at $28^\circ 45^\prime 30^{\prime\prime}$N,
$17^\circ$ $52^\prime$ $48^{\prime\prime}$W and 2250~m above sea level. 
The telescope has a 17-m diameter tessellated parabolic mirror, and is
equipped with a 3.5$^\circ$-3.8$^\circ$ field of view
camera. See~\cite{martinez} for a complete description of the
instrument. 

In this work we show that MAGIC has the capability to contribute to
the growing VHE -ray source catalogue by exploring the part of the
Galactic sources observable from the Northern Hemisphere. The physic
program developed with the MAGIC telescope includes both, topics of
fundamental physics and astrophysics.  In this paper we present the
results regarding the observations of galactic targets. The results
from extragalactic observations are presented elsewhere in these
proceedings~\cite{firpo}. 


\section{The Crab nebula and pulsars} 
\label{sec:crab}
The Crab nebula is a steady emitter at GeV and TeV energies, what
makes it into an excellent calibration candle. This object has
been observed extensively in the past over a wide range of
wavelengths, up to nearly 100 TeV. Nevertheless, some of the
relevant physics phenomena are expected to happen in the VHE domain,
namely the spectrum showing an Inverse Compton (IC) peak close to
100~GeV, a cut-off of the pulsed emission somewhere between 10-100~GeV,
and the verification of the flux stability down to the percent
level. The existing VHE $\gamma$-ray experimental data is well
described by electron acceleration followed by the IC scattering of
photons generated by synchrotron radiation (synchrotron self Compton
process). 

\begin{figure}[h]
  \centering
  \includegraphics[width=7.5cm]{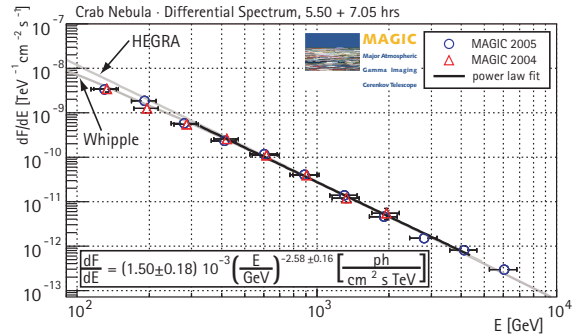}
  \caption{ Energy spectrum above 100 GeV from the Crab nebula
  measured by MAGIC in two different observation seasons.}
  \label{fig:crab}
\end{figure}

Along the first cycle of MAGIC's regular observations, a significant
amount of time has been devoted to observe the Crab nebula, both for
technical and astrophysical studies. A sample of
12~hours of selected data has been used to measure with high precision
the spectrum down to $\sim$100~GeV, as shown in
Figure~\ref{fig:crab}~\cite{wagner}. We have also carried out a search
for pulsed $\gamma$-ray emission from Crab pulsar, albeit with negative results. The derived
upper limits (95\% C.L.) are 2.0$\times 10^{-10}$~ph s$^{-1}$cm$^{-2}$ at 90~GeV and
1.1$\times 10^{-10}$~ph s$^{-1}$cm$^{-2}$ at 150~GeV.

We also carried out a search for pulsed $\gamma$-ray emission from two
milisecond pulsars~\cite{ona} PSR~B1957+20 and PSR~J0218+4232,
albeit without positive result. The corresponding upper limits are
$F_{\mathrm{PSR~B1957+20}} \sim 2.3 \times 10^{-11}$ and 
$F_{\mathrm{PSR~J0218+4232}} \sim 2.9 \times 10^{-11}$ ph s$^{-1}$cm$^{-2}$ for
the steady emission and  
$F_{\mathrm{PSR~B1957+20}} \sim 5.1 \times 10^{-12}$ and 
$F_{\mathrm{PSR~J0218+4232}} \sim 6.5 \times 10^{-12}$ ph s$^{-1}$cm$^{-2}$ for
the pulsed one.


\section{Supernova remnants}
\label{sec:sn}
Shocks produced at supernova explosions are assumed to be the source
of the galactic component of the cosmic ray flux~\cite{zwicky}. The
proof that this is the case could be provided by observations in the
VHE domain. The rationale is that the hadronic component of the cosmic
rays --enhanced close to their source, i.e.\ the SNR-- should produce
VHE $\gamma$-rays by the interaction with nearby dense molecular
clouds. Although recent data seem to indicate that this is the case,
it is difficult to disentangle the VHE component initiated by
hadrons from that produced by Bremsstrahlung and IC processes by
accelerated electrons. Therefore more data in the TeV regime together
with multi-wavelength studies are needed to finally solve the
long-standing puzzle of the origin of galactic cosmic rays.

Within its program of observation of galactic sources, MAGIC has
observed a number of supernova remnants. In particular, we are
observing several of the brightest EGRET sources associated to SNRs,
and the analysis of the acquired data is in progress. On the other
hand, we have confirmed the VHE $\gamma$-ray emission from the SNRs
HESS~J1813-178~\cite{magic_hess1813} and HESS~J1834-087
(W41)~\cite{magic_hess1834}. Our results have confirmed SNRs as a well
established population of VHE $\gamma$-ray emitters. The energy
spectra measured by MAGIC are both well described by an unbroken
power law and an intensity of about 10$\%$ of the Crab nebula
flux. Furthermore, MAGIC has proven its capability to study moderately
extended sources by observing HESS~J1834-087. Interestingly, the maximum of the VHE
emission for this object has been correlated with a maximum in the density of a nearby
molecular cloud. Although the mechanism responsible
for the VHE radiation remains yet to be clarified, this is a hint that
it could be produced by high energy hadrons interacting with the
molecular cloud.

\subsection{Galactic Center}
\label{sec:gc}
We have also measured the VHE $\gamma$-ray flux from the
galactic center~\cite{magic_gc}. 
The possibility to indirectly detect dark matter
through its annihilation into VHE $\gamma$-rays has risen the interest
to observe this region during the last years. Our observations have
confirmed a point-like $\gamma$-ray excess whose location is spatially
consistent with Sgr A* as well as Sgr A East. The energy spectrum of
the detected emission is well described by an unbroken power law of
photon index $\alpha=-2.2$ and intensity about $10\%$ of that of the
Crab nebula flux at 1~TeV. The power law spectrum disfavours dark matter
annihilation as the main origin of the detected flux.
There is no evidence for variability of the flux on hour/day time
scales nor on a year scale. This
suggests that the acceleration takes place in a steady object such as
a SNR or a PWN, and not in the central black hole.



\section{The $\gamma$-ray binary LS~I~+61~303} 
\label{sec:lsi}

\begin{figure}[!t]
\centering
\includegraphics[width=10.5cm]{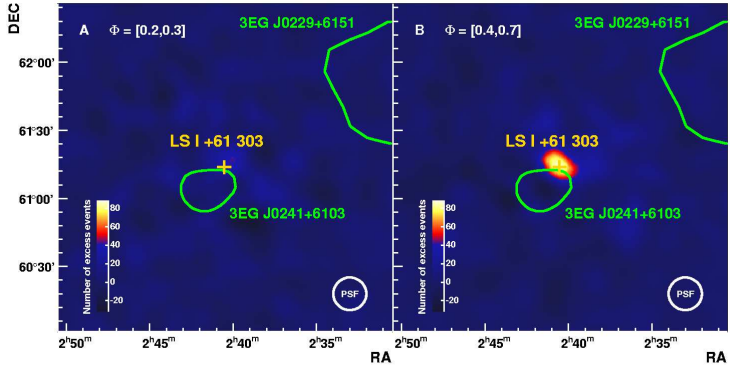}
\caption{
Smoothed maps of $\gamma$-ray excess events above 400 GeV around
LS~I~+61~303 (from \cite{lsi}), for observations around periastron (A) and latter orbital
phases (B).
}
\label{fig:lsi-skymap}
\end{figure}

This $\gamma$-ray binary system is composed of a B0 main sequence star
with a circumstellar disc, i.e. a Be star, located at a distance of
$\sim$2 kpc. A compact object of unknown nature (neutron star or black
hole) is orbiting around it, in a highly eccentric ($e=0.72\pm0.15$)
orbit.
  
LS~I~+61~303 was observed with MAGIC for 54 hours
between October 2005 and March 2006~\cite{lsi}. 
The reconstructed $\gamma$-ray map is shown in
Figure~\ref{fig:lsi-skymap}. The data were first divided into 
two different samples, around periastron passage (0.2-0.3)
and at higher (0.4-0.7) orbital phases. No significant excess in the
number of $\gamma$-ray events is detected around periastron passage,
whereas there is a clear detection (9.4$\sigma$ statistical significance) at
later orbital phases.
Two different scenarios have been involved to explain this high energy
emissions: the microquasar scenario where the $\gamma$-rays are produced
in a radio-emitting jet; or the pulsar binary scenario, where 
they are produced in the shock which is generated by the interaction
of a pulsar wind and the wind of the massive companion.
See~\cite{sidro} for more details.

\paragraph{Acknowledgements.}
We thank the IAC for the excellent working conditions at the
ORM in La Palma. The support of the
German BMBF and MPG, the Italian INFN, the Spanish CICYT is gratefully
acknowledged. This work was also supported by ETH research grant
TH-34/04-3, and the Polish MNiI grant 1P03D01028. 

%
%
%
%
\bibliography{biblio}
\bibliographystyle{unsrt}

%


\printindex
\end{document}